\documentclass[review]{elsarticle}

\usepackage{lineno,hyperref}
\modulolinenumbers[5]
\usepackage{amsmath}

\begin{document}

\begin{frontmatter}

\title{Stationary crack propagation in a two-dimensional visco-elastic network model}
\author{Yuko Aoyanagi}
\author{Ko Okumura\corref{co}}
\cortext[co]{Corresponding author}
\ead{okumura@phys.ocha.ac.jp}

\address{Department of Physics and Soft Matter Center, Ochanomizu University, 2--1--1, 
Otsuka, Bunkyo-ku, Tokyo 112-8610, Japan}

\begin{abstract}
We investigate crack propagation in a simple two-dimensional visco-elastic
model and find a scaling regime in the relation between the propagation
velocity and energy release rate or fracture energy, together with lower and upper 
bounds of the scaling regime. On the basis of our result, the existence of the lower and upper bounds is
expected to be universal or model-independent: 
the present simple simulation model provides generic insight into the physics of crack propagation, 
and the model will be a first step towards the development of a more refined coarse-grained model. 
Relatively abrupt changes of velocity are predicted near the lower and upper bounds for the scaling
regime and the positions of the bounds could be good markers for the development of tough polymers, for
which we provide simple views that could be useful as guiding principles for
toughening polymer-based materials.
\end{abstract}


\end{frontmatter}

\linenumbers

\section{Introduction}
Polymer-based materials are widely used for industrial products and developing
tough polymers are significantly important for our life. Given that material
toughness is governed by cracks at the tips of which stress is concentrated \cite{Anderson,Lawn}, crack propagation in polymer-based materials should be a
subject of wide interest for researchers in academia as well as those in
industry. In fact, fracture energy required for crack propagation and its
dependence on the propagation speed have been studied for various
polymer-based materials, such as adhesive interface \cite{GentPeelingRate1972,Gent1996Langmuir,Chaudhury1999Rate,morishita2008contact}%
, flexible laminates \cite{Kinloch1994peelingRate}, viscoelastic solids \cite{schapery1975theory,GreenwoodJohnsonRate,Langer1989PRA,Persson2005PRE,Tsunoda2000}%
, weakly crosslinked gels \cite{PGGtrumpet,saulnier2004adhesion}, and soft
polymer foam \cite{Kashima2014}. In the case of viscoelastic materials, such
as rubbers and elastomers, a simple scaling regime has been shown
experimentally \cite{Gent1996Langmuir,Tsunoda2000} in the relation between
the fracture energy and velocity when viscoelasticity dominates the fracture
energy (note that rapid crack propagations are strongly affected by inertia \cite{Marder2004PRL,MarderPRL09,Fineberg2005PRL} and that the greatest lower bound for the scaling regime has been discussed in the literature \cite{LakeThomas1967,Hui2003}). This scaling law has been
discussed theoretically using frameworks based on linear viscoelasticity and
linear fracture mechanics by three different groups \cite{GreenwoodJohnsonRate,Langer1989PRA,Persson2005PRE} and, although the
near-crack treatments are different among the groups, they all concluded
essentially the same scaling law in a high velocity limit, suggesting the
importance of the far-field contribution coming from viscoelastic dissipation
occurring at regions remote from crack tips \cite{SoftInterface}.\\
\indent However, the complete physical picture for the far-field viscoelastic regime has
yet to be clarified  with lack of any coarse-grained simulation models for the problem.
We study the crack propagation in a lattice model that incorporates a linear viscoelasticity in a simple manner. The use of lattice model is motivated by the previous theories \cite{GreenwoodJohnsonRate,Langer1989PRA,Persson2005PRE}, in which the dynamics originating from the far-field linear viscoelastic contribution are fairly insensitive to near-crack treatments. As a result, we reproduce crack propagation with a constant velocity. In addition, we find that the velocity as a function of fracture energy or energy release rate exhibits a scaling regime similar to the one discussed in experimental studies \cite{Gent1996Langmuir,Baumberger2006,Lefranc201497}.  Furthermore, we find that there are a lower bound \cite{LakeThomas1967} and an upper bound \cite{PGG1996softAdhesive} for the scaling regime, and we draw simple physical interpretations for the bounds. Since the interpretations are independent of the details of the model, the present simulation model provides generic insight into the physical understanding of the crack propagation, which may be helpful for the development of tough polymer materials.

\section{Simulation model} 
\begin{figure}
\includegraphics[width=1\linewidth]{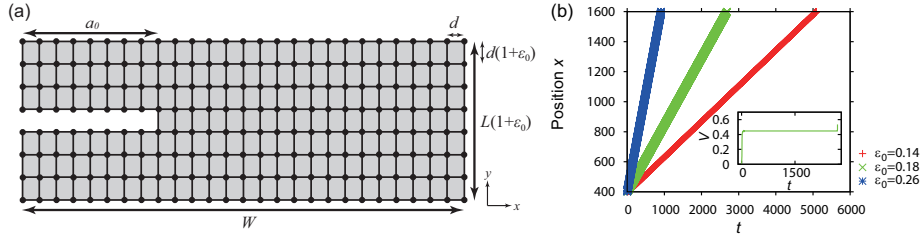}
\caption{(a)Two-dimensional network at the start of simulation. (b) Position of crack tip
$x$ vs time $t$ for three different initial strains $\varepsilon_{0}$ for
$(E,\eta,\varepsilon_{c},d,L)=(100,80,0.35,1,200)$. The inset shows the relation between crack speed $V$
 vs $t$ at the initial strain $\varepsilon_{0} = 0.18$.}%
\label{Fig1}%
\end{figure}

In the simulations performed in the present study, we prepare a two-dimensional square-lattice system with the
lattice constant $d$ as shown in Fig. \ref{Fig1}(a). The width and height are $W$ and $L$, respectively.
Before starting a simulation, we prepare an equilibrium state of the network
with a homogeneous strain $\varepsilon_{0}$ by applying fixed displacements at
the top and bottom of the system. The edge displacements are fixed during the
simulation (fixed-grip condition). The simulation is initiated by introducing
a crack of the initial length $a_{0}$ by cutting  (i.e. removing) the corresponding elastic
bonds as illustrated in Fig. \ref{Fig1}(a). The parameters $(W,a_{0})$ are
fixed to $(1600,400)$ throughout this work in the unit length specified below.

The lattice dynamics is determined by the following mechanism. Each bead in
the lattice feels elastic force from the four nearest neighbors (except for
the beads at the edges), whereas viscous force acts on each bead. Since we
are interested in a purely viscoelastic regime, we neglect the inertia of each
bead. The dynamics of the simulation model can be characterized by the
following equation:%
\begin{equation}
k\Delta x_{i,j}= - \eta\dot{x}_{i,j} \label{e0}%
\end{equation}
where $k$ and $\eta$ are the spring constant and viscosity, respectively. 
Here,  $x_{i,j}$ stands for the vertical position of  the bead located originally at the lattice point ($i, j$)
 and $\dot{x}_{i,j}$ is the time derivative of $x_{i,j}$.
The quantity $\Delta x_{i,j}$ stands for the local displacement, defined as the sum of the elongation of the four bonds (or springs) connecting the lattice point ($i, j$) to the four nearest-neighbor lattice points $x_{i,j}^{(s)} (s=1,..,4)$: 
$\Delta x_{i,j}$ is given by $\Delta x_{i,j} = \sum_{s=1}^4 (x_{i,j}^{(s)}-x_{i,j}-\ell^{(s)})$
with the natural length $\ell^{(s)}$ of the springs with $\ell^{(1)}=\ell^{(3)}=0$ and $\ell^{(2)}=\ell^{(4)}=d$ ($s=1$ and $3$ correspond to shear and $s=2$ and $4$ correspond to stretch) \cite{Aoyanagi2009JPSJ}.
In order for a crack to propagate, every spring is broken
when the force acting on it reaches the critical value $f_{c}$.\\
\indent For later convenience, we define the local ``strain'' and ``stress'' $\varepsilon \equiv \Delta d/d$ and $\sigma \equiv f/d^2 = E \varepsilon$ with the ``elastic modulus'' $E \equiv k/d$. Here, $\Delta d$ is the elongation of a bond and $f$ is the force acting on the bond. Given that there is extensive literature on lattice modelling where relations between lattice parameters and the material Young's modulus and Poisson's ratio are discussed (e.g. \cite{Wang20083459, Jivkov20123089}), it is clear that our results cannot be directly compared with experiment through the ``strain'' and ``stress'' defined above. However, this work discusses fracture mechanical concepts, which are based on continuum theory, and we do not aim at relating our ``stress'' and ``strain'' to measurable macroscopic properties but rather aim at providing physical scenario emerging from a simple model. Accordingly, we use the above definition, which is dimensionally correct and useful to greatly simplify the introduction and discussion of quantities that appear in the fracture mechanical context. With the same spirit, we define $\sigma_c$ by $f_c=\sigma_c d^2$ with $\sigma_c = E \varepsilon_c$ and the principal relaxation time $\tau$ by $\tau =\eta/E$.\\
\indent The units are specified by the fundamental units of length $l_{0}$, elasticity
$E_{0}$, and viscosity $\eta_{0}$, which are all set to one, in the
simulations (for example, the units of time and velocity are given by
$\tau_{0}=\eta_{0}/E_{0}$ and $V_{0}=\ell_{0}/\tau_{0}$, respectively).\\
\indent The creep dynamics of the present model is similar to that of the Kelvin-Voight model:
 under a constant stress, the strain slowly increases with time and finally reaches a constant value.
In fact, the present model possesses $N$ different relaxation times with $N=L/\ell_0$. The details of rheological properties of the simple model will be discussed elsewhere.

\section{Results}

\subsection{Crack propagation with a constant speed}
\begin{figure}
\includegraphics[width=1\linewidth]{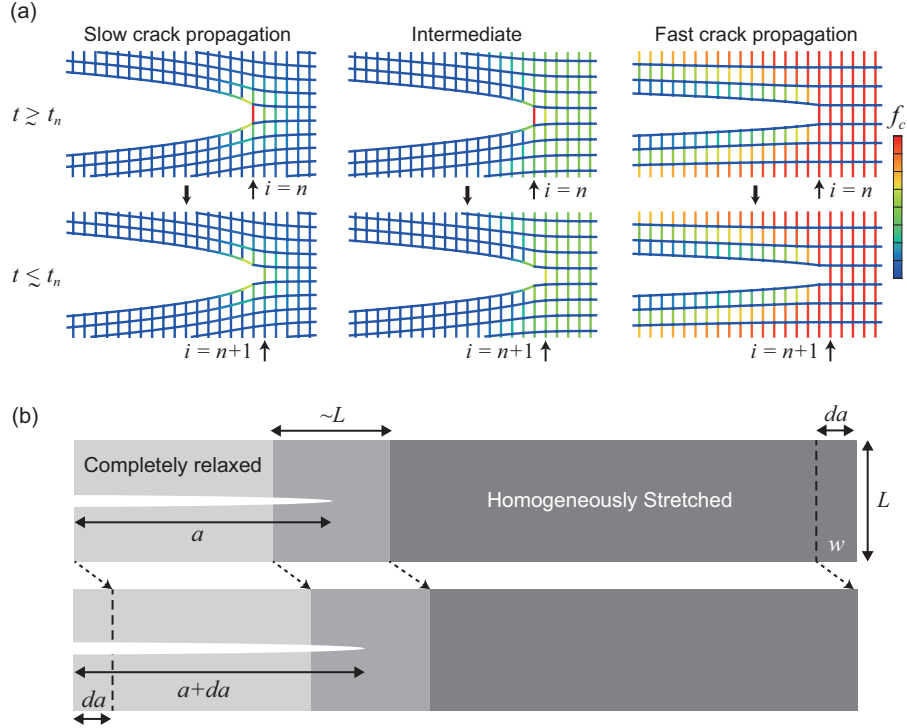}

\caption{ (a) The states
just before and after the bond breaking at the crack tip are shown for slow
crack propagation at $\varepsilon_0=0.070$ (left) and fast propagation at $\varepsilon_0=0.345$ (right), for the same parameter
set as in Fig. {\ref{Fig1}}(b). The middle is obtained at $\varepsilon_0=0.20$, i.e., at a velocity in the middle of the scaling regime shown in Fig. \ref{Fig2} below.  (b) Illustration of the elastic field in the system.}%
\label{Fig1cd}%
\end{figure}

We confirmed that the crack expands with a constant speed for all the
parameters we investigated as demonstrated in Fig. \ref{Fig1}(b). 
As shown in the inset, after a short transient regime, the crack propagation velocity reaches a constant value $V$. 
Figure \ref{Fig1cd}(a) demonstrates
that the crack tip shape changes with speed, which is further discussed in Sec. 4.

\subsection{Fracture energy vs crack propagation speed}
In the present simulations, the energy release rate during the constant-speed
crack propagation is identified with the initially stored elastic energy
multiplied by the system height:%
\begin{equation}
G = wL= \frac{1}{2} E\varepsilon_{0}^{2}L \label{eq1}%
\end{equation}
where $w$ is the density of the initial elastic energy 
$E\varepsilon_{0}^{2}/2$. This is because, as shown in Fig. \ref{Fig1cd}(b), in
the left (right) region away from the crack tip by the distance $\sim$ $L$,
the elastic field is completely relaxed (the elastic field is homogeneous with the initial
energy density $w$) \cite{Lake2003RubberFracture}. Note that $G$ is defined by
$G=-dU/dA$ with $U$ the elastic potential energy and $A$ the fracture surface.
In the present case, $G$ can be interpreted as a velocity-dependent fracture energy since this rate has the meaning of the energy required to create a unit area of fracture surface at a given speed.

\begin{figure}
\includegraphics[width=1\linewidth]{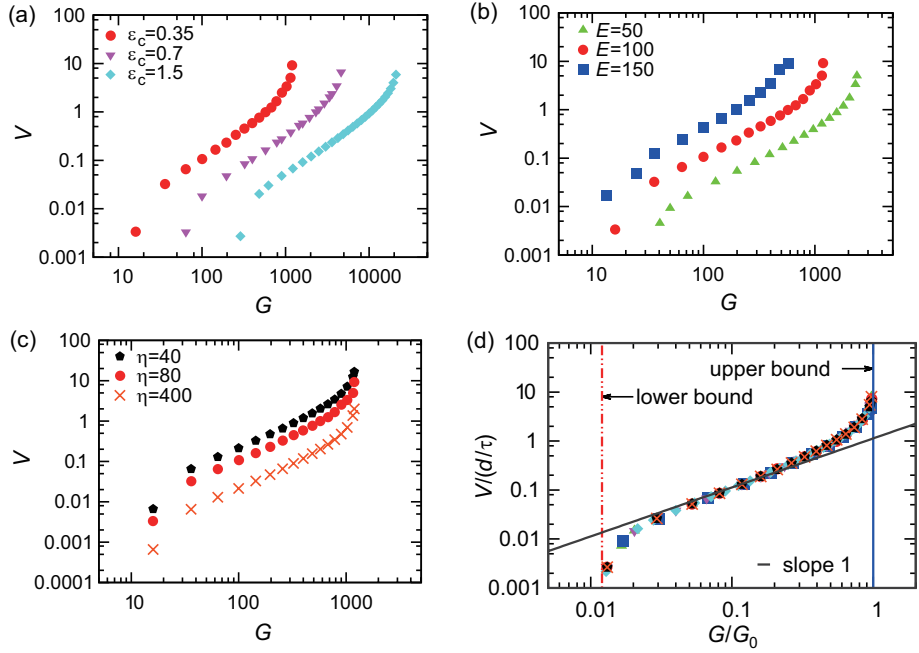}
\caption{Crack propagation
speed $V$ vs. energy release rate $G$ for different values of (a)
$\varepsilon_{c}$ with $(E,\eta)=(100,80)$, (b) $E$ with $(\eta,\varepsilon
_{c})=(80,0.35)$ and (c) $\eta$ with $(E,\varepsilon_{c})=(100,0.35)$, where
$(d,L)$ is fixed to $(1,200)$. (d) $V/(d/\tau)$ vs $G/G_{0}$. The value of $d/L$ in all the cases is $1/200$ and the vertical dashed line indicating the lower bound corresponds to $G/G_0=k_{1} d/L$ with $k_{1}=2.4$. }%
\label{Fig2}%
\end{figure}

\begin{figure}
\includegraphics[width=1\linewidth]{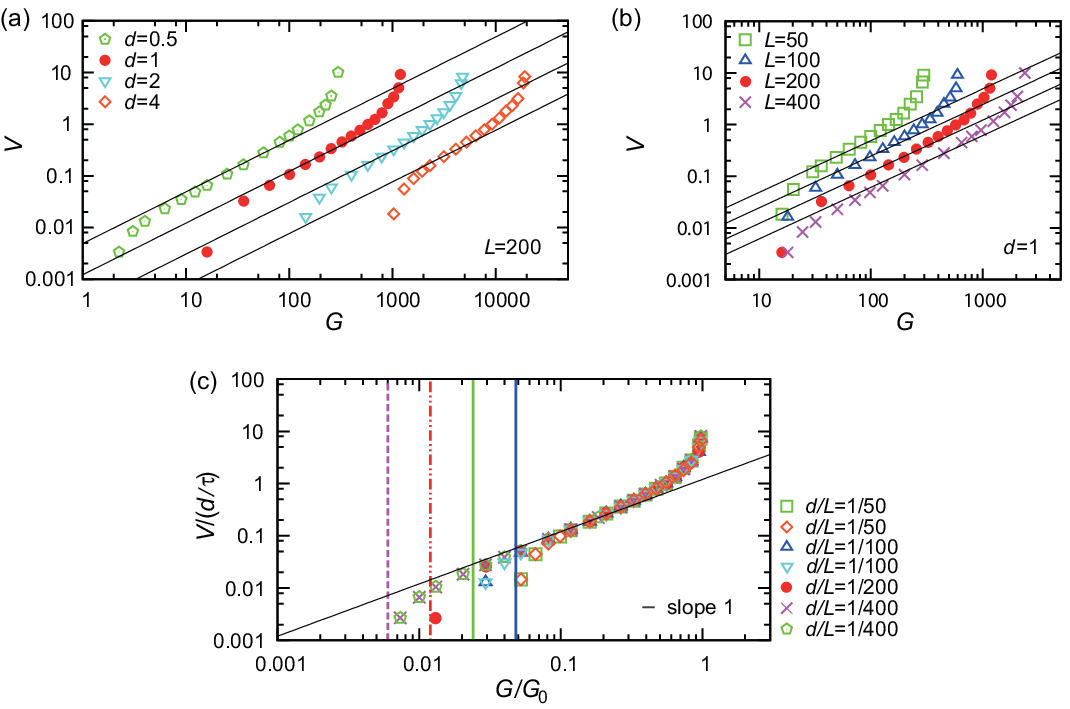}
\caption{Crack propagation
speed $V$ vs. energy release rate $G$ for different values of (a) $d$ with
$L=200$ and (b) $L$ with $d=1$, where $(E,\eta,\varepsilon_{c})=(100,80,0.35)$%
. (c) $V/(d/\tau)$ vs $G/G_{0}$. Four straight lines in (a) and those in (b) correspond to the straight line with slope 1 in (c). In (a) and (b), seven sets of the data with different $(d, L)$ are distinguished by seven different symbols; the data set represented by the filled circles (red circles on the web version) obtained for $d=1$ and $L=200$ is shown in both (a) and (b). These seven sets of data are shown in (c) with using the same symbols. The four vertical lines indicating the lower bounds for four different values of $d/L$ correspond to $G/G_0=k_{1} d/L$ with $k_{1}=2.4$.}%
\label{Fig3}%
\end{figure}

As demonstrated below, the results of simulation show that, when $V/V_0$ is plotted as a function of $G/G_0$, all the simulation data collapse on to a master curve, which can be characterized reasonably well by the following scaling law%
\begin{equation}
\frac{V}{V_{0}} \simeq \left(\frac{G}{G_{0}} \right)^\nu \text{ for } \frac{d}{L}\ll \frac{G}{G_{0}}\ll1 \label{eq4}%
\end{equation}
where the exponent $\nu$ is approximately one, with relatively abrupt changes in velocity at the both ends
($G/G_{0}\simeq d/L$ and $G/G_{0}=1$) of the scaling regime. These abrupt changes imply that the master curve diverges in the upper limit and converges to zero in the lower limit. Here, we have
introduced natural units of the rate $G_{0}$ and the velocity $V_{0}$:%
\begin{equation}
G_{0}=w_{c}L\text{ with }w_{c} =\frac{1}{2} E\varepsilon_{c}^{2} \text{ and }%
V_{0}=\frac{d}{\tau} \label{eq2}%
\end{equation}

In Fig. \ref{Fig2},  the crack propagation speed $V$ is given as a function of
the energy release rate $G$ during the crack propagation for various parameters
$(E,\eta,\varepsilon_{c})$ with fixed $(d,L)$. In Fig. \ref{Fig3}, $V$ is
given as a function of $G$ for various parameters $(d,L)$ with fixed
$(E,\eta,\varepsilon_{c})$. In both cases, when the velocity and the energy
release rate $G$ are renormalized by the natural units $V_{0}$ and $G_{0}$
given in Eq. (\ref{eq2}), all the data are superposed as in Fig. \ref{Fig2}(d)
and Fig. \ref{Fig3}(c), suggesting a linear scaling regime characterized by
the straight line with slope one and the existence of the lower and upper
bounds for the scaling regime with the upper bound $G/G_{0}=1$. We further
confirmed numerically that the lower bound is proportional to $d/L$
in Fig. \ref{Fig3}(c) (this is shown by the fact that the four vertical
lines showing the lower bounds for different $d/L$ are equally spaced), which is theoretically
justified by the arguments in Sec. 4.3.

Some of the exponents in the scaling regions in Figs. \ref{Fig2} and \ref{Fig3} are given numerically as follows. As suggested above, the scaling regime becomes wider as $d/L$ gets smaller, we select from Fig. \ref{Fig3} the data with the smallest to third smallest values of $d/L$ ($d/L=1/400, 1/200$ and $1/100$) and numerically obtained the exponents, which are respectively given as $1.13, 1.23$ and $1.25$. (The exponent is obtained numerically by fitting a straight line to the three points selected in the central region of the scaling regime on a log log plot.) We see that all the values are slightly larger than one and the value gets smaller as $d/L$ becomes smaller. We expect that this effect for finite size of $d/L$ may lead to the result of the exponent one in the small $d/L$ limit as justified in Sec. 4.5, although further confirmation requires a separate study.

\section{Theoretical interpretations}

\subsection{Maximum crack-tip stress on the lattice}
\begin{figure}
\begin{center}
\includegraphics[width=0.5\linewidth]{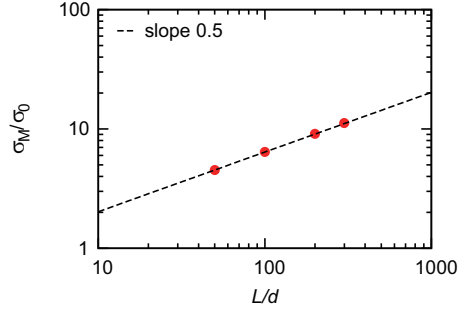}
\end{center}
\caption{The maximum stress at the crack tip in the present lattice model. The simulation data are well on the line with slope 1/2, which confirms Eq. (\ref{eq7}).}%
\label{Fig5}%
\end{figure}

In the static limit, the maximum stress that can appear at the crack tip is
given by
\begin{equation}
\sigma_{M}\simeq\sigma_{0} \left( \frac{L}{d} \right)^{\frac{1}{2}} \label{eq7}%
\end{equation}
This is understood as follows. In the continuum limit, the stress distribution
near the crack tip at the distance $r$ from the tip is generally given by
$\sigma(r)\simeq\sigma_{0}(L/r)^{1/2}$ when the crack size is larger than
$L$ \cite{SoftInterface}. This continuum expression no longer holds when $r$
approaches a critical size below which the system cannot be regarded as a
continuum system anymore. Since this critical scale is given by the lattice
constant $d$, the maximum stress $\sigma_{M}$ that appears at the crack tip
may be given by Eq. (\ref{eq7}). This naively expected relation \cite{Aoyanagi2009JPSJ} is confirmed by
simulation in the present model as shown in Fig. \ref{Fig5}.

\subsection{Mechanism of crack propagation}
As in Fig. \ref{Fig1cd}(a), at the moment $t=t_{n}$, the force acting on the
bond located at the crack tip ($i=n$), reaches the critical value $f_{c}$,
the bond is broken. Just after this moment, the stress on the bond at the new
crack tip ($i=n+1$) has yet to reach the critical value and the stress on
the tip starts to increase till the bond breaks at $f_{c}=\sigma_{c}d^{2}$.
Note that this stress-increasing process is not instantaneous because of the
finite relaxation time $\tau$.

\subsection{Lower and upper bounds for the scaling law}
As seen below, the lower and upper bounds for the scaling law correspond to
the conditions $\sigma_{c}<\sigma_{M}$ and $\sigma_{0}<\sigma_{c}$,
respectively. Equation (\ref{eq7}) implies that, for a given $\sigma_{0}$, a
crack can propagate only if $\sigma_{c}<\sigma_{M}$, from which we obtain
$G/G_{0}=(\sigma_{0}/\sigma_{c})^{2}>(\sigma_{0}/\sigma_{M})^{2} \simeq d/L$:
the lower bound is given by $d/L\simeq G/G_{0}$ as already given in Eq.
(\ref{eq4}). In contrast, in the limit $\sigma_{0}=\sigma_{c}$, we expect the
propagation speed diverges, because any stress concentration is not required
for failure. This leads to the upper bound in Eq. (\ref{eq4}), because
$\sigma_{0}=\sigma_{c}$ can be cast into the form $G/G_{0}=1$.

As indicated in the captions to Figs. \ref{Fig2} and \ref{Fig3}, we find that the lower bounds in all the cases are well described numerically by $G/G_0=k_{1} d/L$ with $k_{1}=2.4$, which is consistent with the results shown in Fig. \ref{Fig5}. The dashed line in Fig. \ref{Fig5} corresponds to $\sigma_M/\sigma_0=k_{2}(L/d)^{1/2}$, i.e., $(\sigma_0/\sigma_M)^2=(k_{2})^{-2}d/L$, with $k_{2}=0.64$. Then, according to the argument in the previous paragraph, the lower bound should be given by $G/G_0=(k_{2})^{-2}d/L$. This means $k_{1}=(k_{2})^{-2}$, which holds well ($2.4$ is nearly equal to $0.64^{-2}$).

\subsection{Change in the shape of crack tip with speed}
The change in the shape of crack tip with propagation speed, as shown in Fig. 
\ref{Fig1cd}(a), is understood as follows. When the propagation is slow, the
shape should become close to a static shape, namely, a parabolic
shape \cite{Anderson,Lawn}. When the propagation is very fast near the limit $\varepsilon = \varepsilon_{c}$, 
the crack shape is practically formed by two parallel lines
(crack surfaces) separated by the distance $\varepsilon_{c}d$ because bonds near the crack tip are broken almost simultaneously without no time for relaxation, which suggests the sharpening of the crack shape at large speeds. Note that when the crack
speed is high, the relaxation of the system continues after the passage of the
crack tip.

\subsection{Interpretation of the scaling regime}
In a scaling regime, if exists, we expect a scaling form $V/V_{0}%
\simeq(G/G_{0})^{\nu}$ with the scaling exponent $\nu$, considering the
natural units $V_{0}$ and $G_{0}$. When the propagation speed is relatively
slow, the released energy per time scaling as $GVd$ is consumed as the
dissipation per time $\eta\dot{\varepsilon}^{2}L^{2}d$. In addition, we may
expect $V$ scales with $\dot{\varepsilon}$, which implies $G\simeq V$, i.e.,
$\nu=1$, in agreement with Eq. (\ref{eq4}).

\section{Discussion}
\subsection{Previous results in accordance with present results}
As shown above, the exponent for the scaling law is reasonably close to one. This might correspond to some experimental observations, see for example \cite{Baumberger2006} or to very specific cases considered theoretically for example in Ref. \cite{Bouchbinder20112279}, while in most of the cases examined in this paper, an exponent $1/2$ is predicted, as observed experimentally in Ref. \cite{Lefranc201497}. Note that different exponents have also been reported for various polymer-based materials (e.g. \cite{Gent1996Langmuir,Tsunoda2000,Persson2005review,Creton2016})

As shown above, we observe that crack shape changes from a parabolic shape at high crack velocities. Our observation is in agreement with previous experiments and simulations, for example, in Refs. \cite{Long201666} and \cite{Morishita2017230}.

\subsection{Effect of inertia}
If we include the inertial effect, the propagation speed $V$ may finally reach
the speed of elastic wave (sound speed $V_{s}$). In such a case, the scaling
regime would end or the second moderate jump (the second region in which relatively abrupt change in velocity is observed) would be cut off (depending on the size of
$\sigma_{c}$) at the corresponding energy release rate $G_{s}$ above which the
propagation speed $V$ is nearly equal to the sound speed $V_{s}$ irrespective
of the value of energy release rate.

\subsection{Lower and upper bounds discussed in previous studies}
Surprisingly, the lower bound for the scaling regime that emerges from the present model turns out to be physically the same with the one discussed in a classic theory, and thus the present model gives novel insight into the classic theory. We showed the lower bound is given by $G/G_0=d/L$, which means that the energy release rate $G$ approaches $E \varepsilon_c^2 d/2$ at the bound. In fact, this expression can be derived from a result of the classic theory by Lake and Thomas \cite{LakeThomas1967} when $d$ is identified with the cross linking distance in the case of rubbers, as suggested in Ref. \cite{LakeThomas1967} with the aide of the result obtained in Ref. \cite{thomas1955rupture}. Thus, the simple physical interpretation of the lower bound given in the present study elucidates an interesting physical meaning of the classic theory: the static fracture energy discussed by Lake and Thomas corresponds to the critical state in which the maximum stress $\sigma_M$ at the crack tip coincides with the intrinsic failure stress $\sigma_c$. 

The upper bound for $G$ is discussed, for example, in Ref. \cite{PGG1996softAdhesive}, by using a model with two characteristic moduli. The present study shows that even a simpler model with a single characteristic modulus possesses the upper bound of different physical origin, which is more fundamental and model-independent. This implies that in a real system the least upper bound could be determined as a result of competition between these two types of upper bounds.

The present simple model suggests physical origins of the existence of two bounds for crack propagation: both bounds originate from stress concentration and the intrinsic failure strength $\sigma_c$. (1) The lower bound is understood from stress concentration as explained in Sec. 4.3 by using Eq. (\ref{eq7}) and Fig. \ref{Fig5} (The maximum stress $\sigma_M$ should be larger than $\sigma_c$). (2) The upper bound $G_0$ is understood from no need for stress concentration by noting Eq. (\ref{eq2}) (At the upper bound, no stress concentration is required for a crack to propagate because the initial strain already reach the critical strain). Since the stress concentration and intrinsic fracture strength are model-independent concepts, our results imply a possibility that the upper and lower bounds could exist in other models from the same physical origins. 

Our results would be useful not only for future fundamental studies but also for future development of tough polymer-based materials. For example, one possible design principle for developing materials highly resistant for crack propagation would be making the value of the lower bound larger; in other words, the lower bound is a good marker for developing tough materials. This is because crack propagation does not occur below the lower bound and, thus, this principle would guide us to reduce the risk of crack propagation in materials. Accordingly, an expression for the lower bound clarifying its dependence on important parameters could be useful and open the possibility for controlling the value of the lower bound, which is a good marker for developing tough materials.

\section*{Acknowledgements}
 The authors thank Dr. Katsuhiko Tsunoda and Yoshihiro Morishita (Bridgestone
Corporation, Japan) for useful discussions. This research was partly supported
by Grant-in-Aid for Scientific Research (A) (No. 24244066) of JSPS, Japan, and
by ImPACT Program of Council for Science, Technology and Innovation (Cabinet
Office, Government of Japan).



\end{document}